# CLASSICAL AND NON-RELATIVISTIC LIMITS OF A LORENTZ-INVARIANT BOHMIAN MODEL FOR A SYSTEM OF SPINLESS PARTICLES.

Sergio Hernández-Zapata[1,3] and Ernesto Hernández-Zapata[2]

[1] Facultad de Ciencias, Universidad Nacional Autónoma de México, Circuito exterior de Ciudad Universitaria, 04510 Distrito Federal, México.
[2] Departamento de Física, Matemáticas e Ingeniería, Universidad de Sonora, C.P. 83600, H. Caborca, Sonora, México.
[3] Author to whom correspondence should be addressed:  shernandezzapata@yahoo.com.mx

## Abstract.

A completely Lorentz-invariant Bohmian model has been proposed recently for the case of a system of non-interacting spinless particles, obeying Klein-Gordon equations. It is based on a multi-temporal formalism and on the idea of treating the squared norm of the wave function as a space-time probability density. The particle's configurations evolve in space-time in terms of a parameter $\sigma$, with dimensions of time. In this work this model is further analyzed and extended to the case of an interaction with an external electromagnetic field. The physical meaning of $\sigma$ is explored. Two special situations are studied in depth: (1) the classical limit, where the Einsteinian Mechanics of Special Relativity is recovered and the parameter $\sigma$ is shown to tend to the particle's proper time; and (2) the non-relativistic limit, where it is obtained a model very similar to the usual non-relativistic Bohmian Mechanics but with the time of the frame of reference replaced by $\sigma$ as the dynamical temporal parameter.

**KEYWORDS:** Bohmian Mechanics; Klein-Gordon equation; Relativistic Quantum Mechanics; multi-temporal formalism; space-time probability density; conditional wave function.

**I.- Introduction.**

Bohmian Mechanics is a non-local deterministic theory of particles in motion that seem to account for all the phenomena of non-relativistic quantum mechanics and solves the problem of the "collapse of the wave function" [1-9]. There are two physical entities in the theory: the wave function and the particles (which do have real trajectories in Bohmian mechanics). In the simple case of a system of spinless particles that do not interact electromagnetically with each other, Bohmian mechanics may be summarized as follows:

The wave function, $\varphi(\vec{x}^{(1)}, \vec{x}^{(2)}, \dots, \vec{x}^{(N)}, t) = R\exp[i\tilde{S}/\hbar]$, for N particles satisfies the usual Schröedinger equation:

$$i\hbar \frac{\partial \varphi}{\partial t} = -\sum_{i=1}^{N} \frac{\hbar^2}{2m_i} \nabla^2_{(i)} \varphi + V\varphi, \qquad (I.1)$$

where $V$ is the potential of interaction with an external electromagnetic field. In Bohm theory, however, equation (I.1) is not the complete description, but it is supplemented by the Bohm equation that describes how the particle velocities are determined by the wave function:

$$\frac{d\vec{X}^{(i)}}{dt} = \frac{1}{m_i} \nabla_{(i)} \tilde{S}(\vec{X}^{(1)}, \vec{X}^{(2)}, \ldots, \vec{X}^{(N)}, t), \quad (I.2)$$

where the index *i* runs over all the particles (*i*=1,...,*N*). As it is common in the bohmian literature, in the previous equations, as in the whole of the paper, the *generic* variables $\vec{x}^{(1)}, \vec{x}^{(2)}, \ldots, \vec{x}^{(N)}$, on which the wave function depends, will be denoted by small letters, while the *configurational* variables $\vec{X}^{(1)}, \vec{X}^{(2)}, \ldots, \vec{X}^{(N)}$ (that is, the actual positions of the particles) will be denoted by capital letters.

J..S. Bell showed that any theory that makes the correct experimental predictions must be non local [2,10], so the explicitly non-local character of Bohmian Mechanics should be not regarded as a defect of the theory. Nevertheless, it does bring the question of its relation to the Special Theory of Relativity. In other words, can a version of Bohm's theory be Lorentz-invariant, in spite of its non-locality? Recently it has been suggested the possibility of reconciling Bohm's theory and Relativity by the use of a multi-temporal formalism [11-18]. In the traditional formalism the wave function of a system of *N* particles depends on *N* spatial variables and a unique time, that is,

$$\vec{x}^{(1)}, \vec{x}^{(2)}, \vec{x}^{(3)}, \ldots, \vec{x}^{(N)}, t.$$

On the other hand, in the multi-temporal formalism the unique time is replaced by a set of *N* different times, one for each particle:

$$\vec{x}^{(1)}, t^{(1)}, \vec{x}^{(2)}, t^{(2)}, \vec{x}^{(3)}, t^{(3)}, \ldots, \vec{x}^{(N)}, t^{(N)} \quad (I.3)$$

This is done with the aim of treating time on an equal footing with space, in the spirit of the Special Theory of Relativity [11-18]. Nikolic [12] has done a very interesting additional contribution by interpreting the squared norm of the wave function $\psi$, in the case of *N* Klein-Gordon equations for *N* particles, as a probability density in a configurational space-time (we will use the symbol $\psi$ for the relativistic wave function to distinguish it from the non-relativistic one, $\varphi$). In [12] these equations, plus the corresponding Bohm-like guide equations, constitute the bohmian model for a system of *N* spinless particles. An important characteristic of the resulting dynamics is that it is a non-local model with the property of being covariant. As the bohmian guide equations are built, an intrinsic parameter $\sigma$ (with dimensions of time) is introduced to parameterize the particle trajectories in space-time (In Ref. [12] the letter s is used to denote this parameter, but we have decided to reserve *S* for the phase of the wave function). Another important property of the dynamics is that of equivariance [5-8]: if, when $\sigma = 0$, the particles are distributed according to the probability density $|\psi|^2$ on the configurational space-time of all particles (this condition is sometimes called "quantum equilibrium" [5-8]), then the particles will remain so distributed for all values of $\sigma$. Furthermore, since $\psi$ is a relativistic scalar the condition of quantum equilibrium is automatically covariant. From our point of view, this is an outstanding characteristic of this model in comparison to other previous models proposed by the same author to describe Klein-Gordon-like systems in a bohmian way [13-18]. In these previous models, where a probability density on space (instead of space-time) is used, the existence of a favored frame of reference where all particles are in "quantum equilibrium" is shown, but this condition is not Lorentz-covariant. For this reason, we believe the model in ref. [12] is a step in an interesting direction. In Ref. [12], however, $\sigma$ is used as a "convenient parameter" that can be forgotten after obtaining the configurational particle world lines. Here there is an essential difference with our point of view. We believe that this parameter may have an interesting physical meaning that we will try to analyze in this work.

The aim of this paper is to extend the Nikolic proposal of a "space-time probability density" to a system where the particles interact with an external electromagnetic field and to study in depth the classical and non-relativistic limits of this bohmian model. The goals that we intend to achieve in this work are the following:

(i) We intend to study in depth the idea of a space-time probability density proposed by Nikolic [12] (an external electromagnetic field is also introduced for a more general and interesting discussion). In order to precise what this means it is necessary to use the parameter σ.
(ii) We want to study the possible physical meaning of the parameter σ itself. Here we will need to study the relevant limits of the Nikolic model.
(iii) We want to study the model under two different situations:

   a) The Planck constant $\hbar$ is much smaller than the characteristic actions of the system. In this case we are dealing with the classical limit and it will be shown that, under the hypothesis of the model (exclusive interaction with an external field, a sort of ideal gas where the particles do not "interact" electromagnetically with each other), the einsteinian mechanics of Special Relativity is wholly recovered. In this limit, after a very natural re-parameterization, σ transforms into the "proper time" of each particle.
   b) The involved velocities are much smaller than the light velocity c. In this case we obtain first a multi-temporal formalism for the particle dynamics but with the special characteristic that the "temporal configuration" can be solved separately. When this "temporal configuration" is substituted into the wave function we obtain a Schrödinger-like equation, but with σ as the temporal parameter, and the corresponding guide equations. This still represents a multi-temporal formalism due to the presence of different constants of integration for each particle. The general relation of this multi-temporal non-relativistic model to the usual non-relativistic Quantum Mechanics will be fully analyzed in further studies. In this work we limited ourselves to show a sufficient (formal) condition (involving low enough precision in the time measurements) that leads to the traditional non-relativistic Bohmian Mechanics (for a system of spinless particles interacting with an external field). The interpretation of the probability density obtained after considering the non-relativistic limit is compared to the interpretation of the probability density in traditional bohmian mechanics.

It is convenient to remark the following: while in the multi-temporal wave function, solution of the Klein-Gordon-like equations, the temporal variables $t^{(1)}$, $t^{(2)}$, ..., $t^{(N)}$ are generic variables completely equivalent to the spatial variables (that is the idea), the variable $t$ of the Schrödinger equation is a configurational variable obtained after substituting into a non-relativistic multi-temporal wave function the temporal configuration (which tell us that that the Schrödinger wave function is a *conditional* wave function [5-8]).

The paper is organized as follows. In section II we present and discuss the bohmian model of the system (spinless particles interacting with an external electromagnetic field), closely related to that of Ref. [12]. In section III we study the classical limit and, working along the lines of Ref. [19], we show that the configuration behaves exactly as predicted by the Einsteinian Mechanics of Special Relativity while the parameter σ tends to the "proper time" of each particle. In section IV we explore the non-relativistic limit and obtain first the non-relativistic multi-temporal model along with the corresponding bohmian guide equations. Then, after solving the temporal configuration in terms of the parameter $\sigma$ and after substituting it into the wave function, we obtain a non-relativistic Bohmian mechanics with a Schrödinger-like *conditional* wave function but with $\sigma$ as the only dynamical temporal parameter [5-8]. We then briefly discuss a sufficient (formal) condition that leads to the usual non-relativistic Bohmian Mechanics for the system of particles in terms of the time $t$ of the frame of reference [1-9] when low enough precision in the time measurements is considered. Finally, in section V we present our conclusions.

**II.- The model.**

We consider a system of $N$ spinless particles. In this article we will use the representation of space-time originally proposed by Minkowski, where there is no difference between covariant and contravariant tensors when thinking on Special Relativity [20]. We associate to each particle a position four-vector $X_\mu^{(i)} \equiv (\vec{X}^{(i)}, icT^{(i)})$, a mass $m_i$ and a charge $e_i$ where the superscript $i$ runs over all the particles ($i=1,...,N$). Just as in the introduction, we use capital letters $(\vec{X}, icT)$ for configurational variables and small letters for the generic ones. The system is under the influence of an external electromagnetic four-potential $A_\mu \equiv (\vec{A}, i\phi)$. Here we will use Gaussian units since they are commonly used in writing wave equations in quantum mechanics.

We will assume that the effective wave function of the system, $\psi = \psi(x_\mu^{(i)})$, is a scalar (that is, it is Lorentz-invariant) and satisfies a set of $N$ (covariant) Klein-Gordon equations of the form:

$$\left(\frac{\hbar}{i}\frac{\partial}{\partial x_\mu^{(i)}} - \frac{e_i}{c}A_\mu^{(i)}\right)\left(\frac{\hbar}{i}\frac{\partial}{\partial x_\mu^{(i)}} - \frac{e_i}{c}A_\mu^{(i)}\right)\psi = -m_i^2 c^2 \psi \qquad (\text{II. 1})$$

where $A_\mu^{(i)}$ is the value of the electromagnetic four-potential in the generic position four-vector corresponding to the *i*-th particle, and where we are using the Einstein summation convention for repeated tensorial indexes.

Writing the wave function in terms of a norm, $R$, and a phase, $S$, in the form $\psi = R exp[i\, S/\hbar]$, we can rewrite each Klein-Gordon equation into two equations as follows:

$$\frac{\partial}{\partial x_\mu^{(i)}}\left(R^2\left(\frac{\partial S}{\partial x_\mu^{(i)}} - \frac{e_i}{c}A_\mu^{(i)}\right)\right) = 0 \qquad (\text{II. 2})$$

$$\left(\frac{\partial S}{\partial x_\mu^{(i)}} - \frac{e_i}{c}A_\mu^{(i)}\right)\left(\frac{\partial S}{\partial x_\mu^{(i)}} - \frac{e_i}{c}A_\mu^{(i)}\right) = -m_i^2 c^2 + \frac{\hbar^2 \Box_{(i)}^2 R}{R} \qquad (\text{II. 3})$$

Where $\Box_{(i)}^2 = \partial/\partial x_\mu^{(i)}\ \partial/\partial x_\mu^{(i)}$ is the D'Alambert operator with respect to the generic coordinates of the i-th particle.

In an analogous manner to what it is usually done in Bohm's theory [1-8], equation (II.2) can be re-interpreted as a (stationary) continuity equation for a probability flux. Here $R^2$ would be interpreted as

*a density of probability in the (configurational) space-time* of all particles in the system, and the field

$$V_\mu^{(i)} = \frac{1}{m_i} \frac{\partial S}{\partial X_\mu^{(i)}} - \frac{e_i}{m_i c} A_\mu^{(i)}$$

can be interpreted as a velocity field four-vector for the *i*-th particle. A quantity $\sigma$, with dimensions of time, is now introduced [12] to parameterize the streamlines given by the velocity field four-vector. Therefore, the following Bohm-like guide equations are proposed [12]:

$$\frac{dX_\mu^{(i)}}{d\sigma} = V_\mu^{(i)} = \frac{1}{m_i} \frac{\partial S}{\partial X_\mu^{(i)}} - \frac{e_i}{m_i c} A_\mu^{(i)} \qquad (\text{II. 4})$$

Note that the parameter $\sigma$ is a scalar since everything else in equation (II.4) is covariant.

The set of N Klein-Gordon equations (II.1) [or, equivalently, II.2 and II.3] together with the set of *N* Bohm-like equations (II.4) constitutes the model for the dynamics of a system of *N* non-interacting spinless particles. Given the wave function, $\psi$, and the values at $\sigma = 0$ of the position four-vector for all particles, $X_\mu^{(i)}$, the equations determine completely its values for every $\sigma$. This is a completely deterministic dynamics, just as in ordinary non-relativistic Bohmian mechanics.

Let us define $\rho(x_\mu^{(i)})$ as the squared norm of the wave function, $\rho(x_\mu^{(i)}) = |\psi(x_\mu^{(i)})|^2$. Just as in non-relativistic Bohmian Mechanics [1-9], the dynamics of the system must be supplemented by a probabilistic assumption concerning initial conditions; namely, that the probability that the actual position four-vectors $X_\mu^{(i)}$ ($\sigma = 0$) are in an interval $\prod_{i,\mu} dx_\mu^{(i)}$ around $x_\mu^{(i)}$ at $\sigma = 0$ is proportional to $\rho(x_\mu^{(i)})$. Thus, as a consequence of equations (II.2-4), the particle position four-vectors will continue to be distributed according to the probability density $\rho(x_\mu^{(i)})$ at other value of $\sigma$. This property of *equivariance* [5-8] follows from the continuity equation (II.2) and the Bohm-like equation (II.4). More precisely, by adding equations (II.2) multiplied by $1/m_i$ and taking into account that $\psi$ does not depend on $\sigma$, we may write following ref. [12] a relativistic conservation equation:

$$\frac{\partial R^2}{\partial \sigma} + \sum_{i=1}^{N} \frac{\partial}{\partial x_\mu^{(i)}} \left( R^2 \left( \frac{1}{m_i} \frac{\partial S}{\partial x_\mu^{(i)}} - \frac{e_i}{m_i c} A_\mu^{(i)} \right) \right) = \frac{\partial R^2}{\partial \sigma} + \sum_{i=1}^{N} \frac{\partial}{\partial x_\mu^{(i)}} \left( R^2 V_\mu^{(i)} \right) = 0,$$

which proves the property of equivariance described above.

Note, however, that we are interpreting $R^2 = |\psi|^2$ as a probability density in the *(configurational) space-time* as a function of $\sigma$ and not as a probability density in the *configurational space* as a function of time [12]. We may call this a *generalized (relativistic) Born rule.*

In Special Relativity the volume of a region of space-time is independent on the frame of reference used for its calculation. Since the jacobian of the Lorentz transformation is equal to unity (here the complex character of time is not important):

$$\begin{cases} x' = \gamma(x - vt) \\ y' = y \\ z' = z \\ ct' = \gamma\left(ct - \dfrac{vx}{c}\right) \end{cases}$$

$$\frac{\partial(x', y', z', ct')}{\partial(x, y, z, ct)} = \begin{vmatrix} \gamma & -\gamma\dfrac{v}{c} \\ -\gamma\dfrac{v}{c} & \gamma \end{vmatrix} = 1,$$

it follows that the volume of a region of space-time is also Lorentz-invariant (this does not mean, however, that $\Delta x'\Delta y'\Delta z'c\Delta t' = \Delta x\Delta y\Delta z c\Delta t$., which is, in general, false). Note also that the density function $\rho(x_\mu^{(i)})$ is a scalar, since the wave function itself $\psi$ is a scalar. The probability assigned by the generalized Born rule (for some value of $\sigma$) to a given region V of the configurational space-time for all particles is given by the integral of $\rho(x_\mu^{(i)})$ over V. It follows that the probability assigned to a certain region V in an inertial frame of reference will be the same than the probability assigned to the Lorentz-transformed region V' in a different frame of reference. In other words, the calculation of the probability is independent on the frame of reference used. Therefore, if the particle position four-vectors are distributed according to the generalized Born rule proposed here in an inertial frame of reference for some value of $\sigma$ *they will be so distributed for all inertial frames of reference and for all values of $\sigma$. The "quantum equilibrium condition" [5-8] is thus, in the model, Lorentz-invariant.*

**III. The classical limit.**

In order to study the classical limit of the model presented in section II we must start by using equation (II.4) to rewrite equation (II.3), obtaining

$$\frac{dX_\mu^{(i)}}{d\sigma}\frac{dX_\mu^{(i)}}{d\sigma} = -c^2 + \frac{\hbar^2 \Box_{(i)}^2 R}{m_i^2 R}, \qquad (\text{III.1})$$

and therefore the squared differential space-time interval may be written as

$$dX_\mu^{(i)} dX_\mu^{(i)} = -c^2 d\tau_{(i)}^2 = \left(-c^2 + \frac{\hbar^2 \Box_{(i)}^2 R}{m_i^2 R}\right) d\sigma^2 \qquad (\text{III.2})$$

where $d\tau_{(i)}$ is the differential of the *i*-th particle proper time. Thus, in the classical limit (where $\hbar$ is much smaller than any characteristic action of the system) we have that

$$-c^2 d\tau_{(i)}^2 = -c^2 d\sigma^2$$

and $$d\tau = d\tau_{(1)} = d\tau_{(2)} = \cdots = d\tau_{(N)}.$$

As the parameter $\sigma$ increases by an amount $d\sigma$ all of the particle's proper times increase by the same amount. Therefore it can be concluded that *the parameter $\sigma$ tends to the particles' proper time*.

An important observation about (III.2) is that if $\hbar^2 \Box_{(i)}^2 R / m_i^2 R > c^2$ the interval is space-like, $d\tau_{(i)}$ is purely imaginary, and the i-th particle leaves the light cone. There is no restriction preventing this in the model, but this does not affect the model Lorentz-invariance. The property of "leaving the light cone in this zone of configurational space" preserves under Lorentz transformations. An analogy with the "tunnel effect" may be considered. When the "quantum potential" term is relevant, the particles may indeed leave the light cone, just as in Non-relativistic Bohmian Mechanics a particle may be in a position where the potential energy is bigger than the total energy. In the classical limit both situations are prohibited. The classical limit implies that the "quantum potential" term becomes irrelevant and the particle's proper time intervals become real, equal to each other, and equal to $d\sigma$.

We are now prepared to discuss the classical limit of the system dynamics. We start by deriving one of the two Klein-Gordon equations [eq. (II.3)], corresponding to the *j*-th particle, with respect to any one of the space-time coordinates for the *i*-th particle, $x_\nu^{(i)}$, leading to

$$2\left(\frac{\partial S}{\partial x_\mu^{(j)}} - \frac{e_j}{c} A_\mu^{(j)}\right) \frac{\partial}{\partial x_\nu^{(i)}} \left(\frac{\partial S}{\partial x_\mu^{(j)}} - \frac{e_j}{c} A_\mu^{(j)}\right) = \frac{\partial}{\partial x_\nu^{(i)}} \left(\frac{\hbar^2 \Box_{(j)}^2 R}{R}\right) \qquad \text{(III. 3)}$$

Using the fact that the external field $A_\mu^{(i)}$ depends only on the space-time coordinates of the *i*-th particle, it is easy to show that

$$\frac{\partial}{\partial x_\nu^{(i)}} \left(\frac{\partial S}{\partial x_\mu^{(j)}} - \frac{e_j}{c} A_\mu^{(j)}\right) = \frac{\partial}{\partial x_\mu^{(j)}} \left(\frac{\partial S}{\partial x_\nu^{(i)}} - \frac{e_i}{c} A_\nu^{(i)}\right) + \frac{e_i}{c} F_{\nu\mu}^{(i)} \delta_{ij} \qquad \text{(III. 4)}$$

where $F_{\nu\mu}^{(i)} = \partial A_\nu^{(i)}/\partial x_\mu^{(i)} - \partial A_\mu^{(i)}/\partial x_\nu^{(i)}$ is the electromagnetic field tensor and $\delta_{ij}$ is the Kronecker delta. Evaluating (III.3) and (III.4) in the actual configurational variables, $x_\mu^{(i)} = X_\mu^{(i)}$, and using equation (II.4) we obtain that

$$m_i V_\mu^{(j)} \frac{\partial V_\nu^{(i)}}{\partial X_\mu^{(j)}} = \frac{e_i}{c} V_\mu^{(i)} F_{\mu\nu}^{(i)} \delta_{ij} + \frac{\partial}{\partial X_\nu^{(i)}} \left(\frac{\hbar^2 \Box_{(j)}^2 R}{2 m_j R}\right) \qquad \text{(III. 5)}$$

Adding together all the *N* equations (III.5), corresponding to the different values of the particle label *j* leads to

$$m_i \frac{d^2 X_\nu^{(i)}}{d\sigma^2} = m_i \frac{dV_\nu^{(i)}}{d\sigma} = m_i \left(\sum_{j=1}^{N} V_\mu^{(j)} \frac{\partial V_\nu^{(i)}}{\partial X_\mu^{(j)}}\right) = \frac{e_i}{c} V_\mu^{(i)} F_{\mu\nu}^{(i)} + \frac{\partial}{\partial X_\nu^{(i)}} \left(\sum_{j=1}^{N} \frac{\hbar^2 \Box_{(j)}^2 R}{2 m_j R}\right) \qquad \text{(III. 6)}$$

In the classical limit (where $\hbar$ is much smaller than any characteristic action of the system) equation (III.6) transforms into *N* decoupled equations. They are identical to the usual equation governing the dynamics of a relativistic particle under the influence of an external electromagnetic field:

$$m_i \frac{d^2 X_\nu^{(i)}}{d\tau^2} = \frac{e_i}{c} \frac{dX_\mu^{(i)}}{d\tau} F_{\mu\nu}^{(i)} \qquad \text{(III. 7)}$$

In equation (III.7) the left-side term is the four-force on the *i*-th particle while the right-side term is the Lorentz force acting upon that particle. These are the equations of the classical Einsteinian Mechanics for a system of $N$ charged particles interacting with an external electromagnetic field but not with each other. It should be remarked that even if the proper time is here a single parameter for all particles we can always re-parameterize in order to get a proper time appropriate to each particle and therefore equations (III.7) are perfectly decoupled.

**IV. The non-relativistic limit.**

Let us now consider the non-relativistic limit of the model. It is convenient to define a new wave phase as:

$$\tilde{S} \equiv S + c^2 \sum_{j=1}^{N} m_j t^{(j)} \qquad (IV.1)$$

It should be noted that, according to this definition, $\tilde{S}$ is not a scalar but depends on the inertial frame of reference.

From the re-definition of the wave phase, Equation (IV.1), it can be proven that, when we change the frame of reference to a primed one that moves with velocity $v$ with respect to the original one, the wave phase transforms in the following way:

$$\tilde{S}' = \tilde{S} - \sum_{j=1}^{N} \left( m_j v x^{(j)} - m_j \frac{v^2}{2} t^{(j)} \right).$$

We start by analyzing how the $N$ Bohm-like equations (II.4) transform when the characteristic velocities are much smaller than $c$. The equation corresponding to the temporal coordinate reads

$$\frac{d(icT^{(i)})}{d\sigma} = \frac{1}{m_i} \frac{\partial S}{\partial (ict^{(i)})} - \frac{e_i}{m_i c} i\phi^{(i)}$$

leading to

$$\frac{dT^{(i)}}{d\sigma} = 1 - \frac{1}{m_i c^2} \frac{\partial \tilde{S}}{\partial t^{(i)}} - \frac{e_i}{m_i c^2} \phi^{(i)} \qquad (IV.2)$$

Thus in the non-relativistic limit we obtain that $dT^{(i)}/d\sigma = 1$ which after integration leads to

$$T^{(i)} = \sigma + \delta_i \qquad (IV.3)$$

where $\delta_i$ is a constant (independent on $\sigma$). It should be remarked that in the non-relativistic limit *the temporal coordinates has decoupled not only from the spatial coordinates but also from the wave function*.

The Bohm-like equations corresponding to the spatial coordinates read

$$\frac{d\vec{X}^{(i)}}{d\sigma} = \frac{1}{m_i}\nabla_{(i)}S - \frac{e_i}{m_i c}\vec{A}^{(i)} = \frac{1}{m_i}\nabla_{(i)}\tilde{S} - \frac{e_i}{m_i c}\vec{A}^{(i)} \qquad (IV.4)$$

In the non-relativistic limit we may neglect all the terms proportional to $c^{-1}$. That is, all the characteristic velocities of the system are considered to be much smaller than the velocity of light $c$. Therefore, equations (IV. 4) transform, in the non-relativistic limit, into the following equation:

$$\frac{d\vec{X}^{(i)}}{d\sigma} = \frac{1}{m_i}\nabla_{(i)}\tilde{S} \qquad T^{(i)} = \sigma + \delta_i \qquad (IV.5)$$

Let us now consider the Klein-Gordon equations (II.2-3) that describe the evolution of the wave function. Equation (II.3) can be rewritten in terms of $\tilde{S}$ in the following way:

$$\left(\nabla_{(i)}\tilde{S} - \frac{e_i}{c}\vec{A}^{(i)}\right)\cdot\left(\nabla_{(i)}\tilde{S} - \frac{e_i}{c}\vec{A}^{(i)}\right) + 2m_i\left(\frac{\partial \tilde{S}}{\partial t^{(i)}} + e_i\phi^{(i)}\right) - \frac{1}{c^2}\left(\frac{\partial \tilde{S}}{\partial t^{(i)}} + e_i\phi^{(i)}\right)^2 = \frac{\hbar^2 \Box_{(i)}^2 R}{R} \qquad (IV.6)$$

Here we have used the identity: $\partial S/\partial(ict^{(i)}) = im_i c - (i/c)\partial\tilde{S}/\partial t^{(i)}$.

Analogously, equation (II.2) can be rewritten in terms of $\tilde{S}$ as

$$\nabla_{(i)}\cdot\left(R^2\left(\nabla_{(i)}\tilde{S} - \frac{e_i}{c}\vec{A}^{(i)}\right)\right) + \frac{\partial}{\partial t^{(i)}}\left(R^2\left(m_i - \frac{1}{c^2}\frac{\partial \tilde{S}}{\partial t^{(i)}} - \frac{e_i}{c^2}\phi^{(i)}\right)\right) = 0 \qquad (IV.7)$$

In the non-relativistic limit we may neglect all the terms proportional either to $c^{-1}$ or to $c^{-2}$.

Therefore, equations (IV. 6-7) transform into

$$\nabla_{(i)}\tilde{S}\cdot\nabla_{(i)}\tilde{S} + 2m_i\left(\frac{\partial \tilde{S}}{\partial t^{(i)}} + e_i\phi^{(i)}\right) = \frac{\hbar^2\nabla_{(i)}^2 R}{R} \qquad (IV.8)$$

and

$$\nabla_{(i)}\cdot\left(\frac{R^2}{m_i}\nabla_{(i)}\tilde{S}\right) + \frac{\partial}{\partial t^{(i)}}(R^2) = 0 \qquad (IV.9)$$

It is easy to see that equations (IV.8-9) are the imaginary and real parts of the following Schröedinger-like equations (one for each particle):

$$i\hbar\frac{\partial \psi}{\partial t^{(i)}} = -\frac{\hbar^2}{2m_i}\nabla_{(i)}^2\psi + e_i\phi^{(i)}\psi \qquad (IV.10)$$

if we redefine the effective wave function in the non-relativistic case as $\psi = R\exp[i(\tilde{S}/\hbar)]$, with the new phase $\tilde{S}$ replacing the old phase $S$.

Equations (IV.5) and (IV.10) describe a dynamics that is already non-relativistic but it is still based on a multi-temporal formalism.

Since the evolution in $\sigma$ of the temporal coordinates is already known [Equation (IV.3)], we may define a new *conditional* wave function $\varphi(\vec{x}^{(1)}, \vec{x}^{(2)}, \ldots, \vec{x}^{(N)}, \sigma) \equiv \psi(\vec{x}^{(1)}, \sigma + \delta_1, \ldots, \vec{x}^{(N)}, \sigma + \delta_N)$ by evaluating the original wave function in the *actual* values of the temporal coordinates. By adding the $N$ equations (IV.10) for all particles and using the chain rule of Differential Calculus, we obtain that

$$i\hbar \frac{\partial \varphi}{\partial \sigma} = i\hbar \sum_{i=1}^{N} \frac{\partial \psi}{\partial t^{(i)}} = -\sum_{i=1}^{N} \frac{\hbar^2}{2m_i} \nabla_{(i)}^2 \varphi + V\varphi \qquad (\text{IV}.11)$$

Here the potential energy have been defined as $V = \sum_{i=1}^{N} e_i \phi^{(i)}$. We have now an equation very similar to that of Schrödinger, but with the intrinsic time $\sigma$ as the only dynamical temporal parameter.

In usual Bohmian Mechanics, when $M - N$ of the spatial coordinates of a system of $M$ particles are solved, $\varphi(\vec{x}^{(1)}, \vec{x}^{(2)}, \ldots, \vec{x}^{(N)}, t) = \psi(\vec{x}^{(1)}, \vec{x}^{(2)}, \ldots, \vec{x}^{(N)}, \vec{X}^{(N+1)}(t), \ldots, \vec{X}^{(M)}(t), t)$ is now a conditional wave function [5-8] that guide the dynamics of the subsystem of $N$ particles. The corresponding probability of the configuration in the subsystem of $N$ particles is [5-8]

$$\left|\varphi(\vec{x}^{(1)}, \vec{x}^{(2)}, \ldots, \vec{x}^{(N)}, t)\right|^2 d^3\vec{x}^{(1)} d^3\vec{x}^{(2)} \ldots d^3\vec{x}^{(N)}$$

Analogously, here where we have solved the dynamics for the temporal coordinates, in the non-relativistic case, the resulting conditional wave function gives the probability in the usual space in the following form:

$$\left|\varphi(\vec{x}^{(1)}, \vec{x}^{(2)}, \ldots, \vec{x}^{(N)}, \sigma)\right|^2 dV^{(1)} dV^{(2)} \ldots dV^{(N)} \qquad (\text{IV}.12)$$

The dynamics of the system in the non-relativistic limit, analyzed in this section, is, thus, the Schrödinger-like equation (IV.11) plus the Bohm-like equation (IV.5), where $\tilde{S}$ is the phase of the (conditional) wave function

$$\varphi(\vec{x}^{(1)}, \vec{x}^{(2)}, \ldots, \vec{x}^{(N)}, \sigma) = \psi(\vec{x}^{(1)}, \vec{x}^{(2)}, \ldots, \vec{x}^{(N)}, \sigma + \delta_1, \sigma + \delta_2, \ldots, \sigma + \delta_N), \qquad (\text{IV}.13)$$

evaluated at the configurational variables $\vec{X}^{(j)}$ and in the expressions (IV.3). Formally we have obtained the same equations than in the usual non-relativistic Bohmian Mechanics (Equations (I.1) and (I.2)) but the temporal parameter $\sigma$ does not coincide with "the time $t$ of the frame of reference" used in the traditional non-relativistic quantum-mechanical description.

The non-relativistic Bohmian model given by equations (IV.5) and (IV.11) is still multi-temporal due to the fact that we have different values of the integration constants $\delta_i$ for each particle. This is a consequence of the multi-temporal character of the original relativistic model. The relation of this model to the usual non-relativistic Quantum Mechanics will be fully analyzed in further studies. Here we will just briefly discuss a sufficient (formal) condition that leads to the traditional uni-temporal non-relativistic Bohmian model. We define

$$\Lambda = max_{i,j=1,\ldots,N} |\delta_i - \delta_j|.$$

Since the system of particles is finite, $\Lambda$ should also be finite but otherwise may have any value. Let us now assume that we measure the time with precision low enough so that the measurement uncertainty $\varepsilon$ is bigger than $\Lambda$, in such a way that the clocks cannot resolve the differences in the values of the constants $\delta_i$ and thus, at this level of precision,

$$\delta_1 = \delta_2 = \cdots = \delta_N = \delta = constant \qquad (IV.14)$$

Substituting this result into Equations (IV.13), (IV.5), and (IV.11), we obtain respectively the wave function, the Bohm guide equation, and the Schrödinger equation in the traditional non-relativistic Bohmian analysis, where the intrinsic time $\sigma$ is replaced by the time $t$ of the frame of reference ($t = \sigma + \delta$).

The probability of finding the $i$-th particle in a volume $dV^{(i)}$ ($i = 1, \ldots, N$), given by equation (IV.12), under the restriction of low precision in time measurements, transforms into a probability proportional to

$$\left|\varphi\left(\vec{x}^{(1)}, \vec{x}^{(2)}, \ldots, \vec{x}^{(N)}, t\right)\right|^2 dV^{(1)} dV^{(2)} \ldots dV^{(N)}. \qquad (IV.15)$$

This non-relativistic probability is inherited from the initial assumption that the $N$ particles are in quantum equilibrium in the relativistic model in terms of the parameter $\sigma$. In other words, we do not have to impose the condition that the initial distribution at time $t = t_0$ is given by (IV.15) (that is, the usual Bohmian quantum equilibrium assumption), since this is implied by the relativistic quantum equilibrium assumption. The stated low-precision condition on the time measurements is thus a sufficient condition to derive the usual uni-temporal non-relativistic Bohmian Mechanics. Since the typical values of the constants $\delta_i$ are unknown, further studies are needed in order to investigate if the low-precision condition is valid for the typical experimental situations in non-relativistic Quantum Mechanics.

**V. Conclusions.**

We have described and analyzed a covariant model very similar to that of Reference [12] in order to show some of the general properties we will obtain when extending Bohm's ideas to Special Relativity. We believe that, using the idea of a "space-time probability density" as in [12], many Bohm-like covariant models may be designed, even for particles with a spin different than zero and that do interact electromagnetically with each other. We intend to explore further this idea in our future work. The criterion of covariance for the models may be used to choose between different Bohmian alternatives describing the same sort of particles. We have extended the model proposed by Nikolic in [12] to include an interaction with an external electromagnetic field and we have studied the behavior of the model under two different contexts: the classical and the non-relativistic limits. In the classical limit we have recovered the Einsteinian Dynamics for a system of particles interacting with an external electromagnetic field. In the non-relativistic limit, we have derived the same equations than in the traditional Bohmian Mechanics but with $\sigma$ as the dynamical temporal parameter. We have also suggested that, under the restriction of low precision in time measurements, this model leads to the traditional Bohmian Mechanics.

*We would like to dedicate this work to our teacher Dr. Luis de la Peña Auerbach, who has always promoted the exploration of alternatives to the single orthodoxy in Quantum Mechanics.*

REFERENCES

[1] Bohm, D.: A suggested interpretation of the quantum theory in terms of "hidden variables", Parts. 1 and 2. *Phys. Rev*. **89**, 166-193 (1952).

[2] Bell, J.S.: Speakable and Unspeakable in Quantum Mechanics. Cambridge University Press, Cambridge (1993).

[3] Durr, D., Teufel, S.: Bohmian Mechanics. The Physics and Mathematics of Quantum Theory. Springer-Verlag, Berlin Heidelberg (2009).

[4] Berndl, K., Daumer, M., Durr, D., Goldstein, S., Zanghi, N.: A Survey of Bohmian Mechanics. *Il Nuovo Cimento* **110B**, 737-750 (1995).

[5] Durr, D., Goldstein, S., Zanghi, N.: Quantum Mechanics, Randomness, and Deterministic Reality. *Physics Letters A* **172**, 6-12 (1992).

[6] Durr, D., Goldstein, S., Zanghi, N.: A Global Equilibrium as the Foundation for Quantum Randomness. *Foundations of Physics* **23**, 721-738 (1993).

[7] Durr, D., Goldstein S., Zanghi, N.: Bohmian Mechanics and Quantum Equilibrium. In: Albeverio,S., Cattaneo, U., Merlini, D. (eds.): Stochastic Processes, Physics and Geometry II, pp. 221-232. World Scientific, Singapore (1995).

[8] Durr, D., Goldstein S., Zanghi, N.: Quantum equilibrium and the role of operators as observables in Quantum Theory. *Journal of Statistical Physics* **116**: 959-1055 (2004).

[9] Tumulka, R.: Understanding Bohmian Mechanics: a dialogue. *American Journal of Physics* **72** (9): 1220-1226 (2004).

[10] Bricmont, J.: Http://www.fyma.ucl.ac.be/files/meaningWF.pdf. Cited 8 Oct 2009 (2009).

[11] Berndl, K., Durr, D., Goldstein, S., Zanghi, N.: Nonlocality, Lorentz invariance, and Bohmian quantum theory. *Physical Review* A **53** (4): 2062-2073 (1996).

[12] Nikolic, H.: Time in Relativistic and Nonrelativistic Quantum Mechanics. *International Journal of Quantum Information* **7** (3): 595-602 (2009).

[13] Nikolic, H.: Relativistic quantum mechanics and the Bohmian interpretation. *Foundations of Physics Letters* **18** (6): 549-561 (2005).

[14] Nikolic, H.: Covariant many-fingered time Bohmian interpretation of quantum field theory. *Physics Letters A* **348** (3-6): 166-171 (2006).

[15] Nikolic, H.: Relativistic Bohmian interpretation of quantum mechanics. Conference on the present status of Quantum Mechanics. *AIP Conference Proceedings* **844**: 272-280 (2006).

[16] Nikolic, H.: Quantum mechanics: myths and facts. *Foundations of Physics* **37** (11): 1563-1611 (2007).

[17] Nikolic, H.: Probability in relativistic quantum mechanics and foliation of spacetime. *International Journal of Modern Physics A* **22** (32): 6243-6251 (2007).

[18] Nikolic, H.: Probability in relativistic Bohmian mechanics of particles and strings. *Foundations of Physics* **38** (9): 869-881 (2008).

[19] Allori, V., Durr, D., Goldstein, S., Zanghi, N.: Seven Steps towards the Classical World. Journal of Optics B **4**, 482-488 (2002).

[20] Einstein, A.: Grundgedancen und Methoden der Relativitätstheorie, in ihrer Entwivehung darqestellt. In: The Collected Papers of Albert Einstein. Princeton University Press, Princeton (2002).